\documentclass{aa}

\usepackage{graphicx}
\usepackage{txfonts}
\usepackage{bm}
\begin{document}

   \title{Failed ejection and oscillations of a current-carrying filament\\ balanced by gravity}
   \titlerunning{Failed ejection of filament balanced by gravity}

   \author{P. Jel\'\i nek
          \inst{1},
          M. Karlick\'y
          \inst{2}
          \and
          S. Belov
          \inst{1}
          }

   \institute{University of South Bohemia, Faculty of Science, Department of Physics, Brani\v sovsk\'a 1760, CZ -- 370 05 \v{C}esk\'e
             Bud\v{e}jovice, Czech Republic\\
              \email{pjelinek@prf.jcu.cz}
         \and
             Astronomical Institute of the Czech Academy of Sciences, Fri\v{c}ova 258, CZ -- 251 65 Ond\v rejov, Czech
             Republic\\
             }

   \date{Received ; accepted }

  \abstract
   {Solar filaments are often associated with solar eruptions and coronal mass ejections. However, in some cases, the ejection process is halted, resulting in a failed eruption. Understanding the processes that occur after filament destabilization is therefore of great importance.}
   {In this study, we investigate the post-destabilization evolution of a filament in a gravity-balanced model.}
   {We adopt the filament model proposed by Solov'ev (2010), in which a dense filament is supported against gravity by the repulsive force between the filament current and its sub-photospheric image. We first performed an analytical investigation of this model. For the numerical study, we use a two-dimensional magnetohydrodynamic (MHD) model that solves the MHD equations with the \texttt{Lare2d} numerical code.}
  {In this filament model, analytical expressions are derived for the electric current density, plasma density, and their spatial distributions as functions of the model parameters. The total electric current and the filament weight are also calculated.
For the numerical simulations, we constructed an equilibrium filament characterized by a magnetic field of $B_0 = 10^{-3}~\mathrm{T}$, mass density $\rho_0 \sim 1.3 \times 10^{-9}~\mathrm{kg\,m^{-3}}$, and temperature $T \sim 13000~\mathrm{K}$. The system was destabilized either by increasing the currents or by reducing the filament density, and its evolution was computed. In both destabilization regimes, the filament was ejected, then halted at a certain altitude, and subsequently fell back, repeating this cycle with a period of about 600 s. The maximum filament ejection velocity was approximately $80$ and $40~\mathrm{km\,s^{-1}}$, respectively.
Beneath the ejected filament a current sheet forms, where magnetic reconnection occurs.
The maximum ejection altitudes were determined as functions of both the destabilizing currents and the degree of filament plasma dilution. Finally, we compared results of this MHD model with those of an ideal vacuum model and discussed all results.}
{}

   \keywords{Sun: filaments -- Sun: flares -- Methods: numerical -- Magnetohydrodynamics (MHD)}

   \maketitle
%
\section{Introduction}
Solar filaments, also known as prominences, are dense and cool plasma structures (particle concentration $10^{10} - 10^{12}~\mathrm{cm^{-3}}$ and temperature in the coldest part ranging from $5000$ to $20000~\mathrm{K}$), suspended in the solar corona \citep{1995ASSL..199.....T,2019A&A...631A.146S,2024AstBu..79..508S}. They are often associated with solar eruptions and coronal mass ejections (CMEs). 
There are several models describing these filaments (see, e.g., \cite{2014masu.book.....P}). Among them, the inverse-polarity model appears to be the most significant. The first such model, proposed by \cite{1974A&A....31..189K}, considered a circular filament (prominence) carrying an electric current that forms at a certain height within a vertical current sheet. In this model, the dense filament is supported against gravity by the upward Lorentz (repulsive) force between the filament current and its sub-photospheric image current. This model was later extended by including the effect of an external coronal magnetic field \citep{1978SoPh...59..115V}. In this extended model, and within the so-called strong magnetic field approximation, \cite{1991ApJ...373..294F} neglected gravity and investigated the mechanism of filament ejection.
Later, a model similar to that of \cite{1974A&A....31..189K} was proposed by \cite{2010ARep...54...86S}.

On the other hand, \cite{1999A&A...351..707T} developed a model in which the repulsive force between the filament current and its mirror current is balanced by a surrounding potential magnetic field. This model has been widely and successfully used in numerical simulations of solar flares and CMEs; see, for example, \cite{2024ApJ...962..149T,2024MNRAS.529..761L}. However, a key limitation of these simulations is the assumption of a zero plasma beta, which neglects the effects of gravity.

Such simplifications, although computationally efficient, may lead to inaccuracies when modelling filament stability under realistic coronal conditions, where finite plasma pressure and gravitational stratification play significant roles \citep{2005ApJ...630..543F,2013ApJ...766..126H}. Moreover, recent observations suggest that the mass loading and thermal structuring of filaments may crucially affect their evolution and eventual eruption \citep{2018LRSP...15....7G}. These factors can influence the threshold for instability, as well as the nonlinear dynamics of the eruption process.
Filament eruptions are often interpreted in terms of ideal MHD instabilities, such as the torus and kink instabilities, which depend on the geometry of the magnetic field and current distribution \citep{2006PhRvL..96y5002K,2010ApJ...708..314A}. The onset of these instabilities may be triggered either by slow quasi-static evolution due to photospheric motions or by impulsive external perturbations. Therefore, developing models that include gravitational forces, finite plasma beta, and realistic current configurations is essential for understanding the pre-eruption equilibrium and the conditions under which filaments lose stability.
Observationally, filaments appear as dark, elongated structures in H-alpha or EUV images when viewed against the solar disk, and as bright prominences above the limb. Their internal structure reveals fine-scale threads aligned with the local magnetic field, indicating a strong coupling between thermodynamic and magnetic properties. However, the precise magnetic topology of filaments remains difficult to determine directly due to limitations in coronal magnetic field measurements \citep{2010SSRv..151..333M,2014LRSP...11....1P}. As a result, theoretical and numerical models remain crucial for interpreting observational data and for exploring mechanisms of filament formation and destabilization.
The long-term stability of filaments also depends on their anchoring in the chromosphere and on the large-scale magnetic environment of the corona. In particular, the overlying magnetic arcade can provide stabilizing tension, and its gradual evolution can lead to the loss of equilibrium. Models that can account for these multi-scale processes, including magnetic reconnection and current dissipation, are necessary to fully capture the physics of filament eruptions.

In this paper, we consider the Solov’ev filament model \citep{2010ARep...54...86S}, in which the repulsive force between the filament current and its mirror current is balanced by gravity. Our objective is to investigate what occurs after a filament in this gravity-balanced configuration becomes destabilized.

The paper is structured as follows: Section 2 introduces the filament model and its analytical analysis. Section 3 describes the corresponding numerical MHD model. Section 4 presents the results. Finally, Section 5 discusses the results in comparison with those of the ideal vacuum model presented in the Appendix, and Section 6 presents the conclusions.

\section{Filament model}
We use the two-dimensional (2D) filament model proposed by \cite{2010ARep...54...86S,2012Ge&Ae..52.1062S}.
In this model, the filament, carrying an electric current, is in equilibrium due to its own weight within the gravitationally stratified solar atmosphere.
The magnetic field components $B_x$ and $B_y$ in the plane intersecting the filament are expressed via the vector potential $\mathbf{A}$, ensuring that the solenoidal condition, which requires the divergence of the magnetic field to be zero, is automatically satisfied
\begin{equation}
A(x,y)=B_0 \frac{y - h_0}{ 1 +k^2 x^2 + k^2 (y - h_0)^2},
\label{eq1}
\end{equation}
as:
\begin{equation}
B_x = B_\mathrm{0} \frac{1 + k^2 x^2 - k^2 (y - h_0)^2}{[1 + k^2 x^2 + k^2(y - h_0)^2]^2},
\label{eq2}
\end{equation}
and
\begin{equation}
B_y = B_\mathrm{0} \frac{2 k^2 x (y - h_0)}{[1 + k^2 x^2 + k^2(y - h_0)^2]^2},
\label{eq3}
\end{equation}
where $x$ and $y$ are the spatial coordinates, $k$ is a free parameter, which defines the scale of the system, $B_\mathrm{0}$ is the magnetic field at $h_\mathrm{0}$, which is the height of the center of the double-structure of the filament (the center between the coronal and sub-photospheric parts).
Analyzing the Eq.~(\ref{eq1}), the locations of the extremes of the vector potential $A(x,y)$ are in
\begin{equation}
x_{\rm A} = 0, \quad y_{\rm A}=h_0 \pm \frac{1}{k}.
\label{eq4}
\end{equation}

Similarly, analyzing Eq.~(\ref{eq2}), the locations of the extremes of the magnetic field component $B_x$ are at
\begin{equation}
x_{\rm Bx} = 0, \quad y_{\rm Bx}=h_0 \pm \frac{\sqrt{3}}{k},
\label{eq5}
\end{equation}
and the locations of $B_x$ = 0 are at
\begin{equation}
x_{\rm Bx0} = 0, \quad y_{\rm Bx0}=h_0 \pm \frac{1}{k}.
\label{eq6}
\end{equation}

On the other hand, the magnetic field component $B_\mathrm{z}$, which is perpendicular to the $x-y$ plane, is expressed as \citep{2020A&A...637A..42J}
\begin{equation}
	B_z^2(x,y) = B_{ex}^2 -B_0^2\frac{\alpha^2k^2(y-h_0)^2}{[1+k^2x^2+k^2(y-h_0)^2]^2}, 
\label{eq7}
\end{equation}
where $\alpha$ is an arbitrary positive constant and $B_\mathrm{ex}$ is an external magnetic field at infinity. The parameter $\alpha$ determines the twist of the magnetic field lines in the filament. 

Finally, the mass density and pressure distribution in the model are \citep{2020A&A...637A..42J}
\begin{equation}
\varrho(x,y) = \varrho_h(y) +\frac{B_0^2}{2 \mu_0 g_\odot} \frac{8 k^2 (y-h_0)}{[1 + k^2x^2 +k^2(y-h_0)^2]^4},  
\label{eq8}
\end{equation}

\begin{equation}
 p(A,y) = p_h(y) - \frac{B_z^2(A)-B_z^2(0)}{2\mu_0} + \frac{4 A^4 k^2}{2\mu_0 B_0^2 (y - h_0)^2},  
\label{eq9}
\end{equation}
where $\mu_0 = 1.26 \times 10^{-6}~\mathrm{H m}^{-1}$ is the vacuum magnetic permeability, $g_\odot = 274~\mathrm{m s^{-2}}$ is the gravitational acceleration at the Sun. Here, $\varrho_h(y)$ is the mass density expressed as, see, e.g. \citep{2014masu.book.....P}
\begin{equation}
\varrho_h(y) = \varrho_0 \frac{T_\mathrm{0}}{T(y)}\exp\left({-\int_0^h}\frac{\mathrm{d}y}{H(y)}\right),
\label{eq10}
\end{equation}
and $p_h(y)$ is a pressure in the gravitationally stratified atmosphere without the filament
\begin{equation}
 p_h(y) = p_0\exp\left({-\int_0^h}\frac{\mathrm{d}y}{H(y)}\right),
\label{eq11}
\end{equation}
where $\varrho_0$, $T_0$ and $p_0$ are the base mass density, temperature, and pressure (at photosphere), respectively. The pressure scale height $H(y)$ is determined by the temperature of the plasma and the local gravitational acceleration, and is given by the formula
\begin{equation}
H(y) = \frac{k_B T(y)}{\tilde{\mu} m_p g_\odot},
\end{equation}
\label{eq12}
where $\tilde{\mu}$ is the mean particle mass and the temperature $T(y)$ is defined by Eq.~(\ref{eq20}).

Using Eqs. (\ref{eq2}), (\ref{eq3}) and the expression for the electric current density $\bm{\mathrm{j}}=(\nabla \times \bm{\mathrm{B}})/\mu_0$ the $z$-component of this density can be expressed as 
\begin{equation}
j_z = \frac{8 B_0}{\mu_0} \frac{k^2(y-h_0)}{[1+k^2x^2+k^2(y-h_0)^2]^3}. 
\label{eq13}
\end{equation}

Analyzing this relation, the locations of the maximal absolute value of the current density $|j_z|_{\rm max}$ are at
\begin{equation}
x_{\rm |j_z|max} = 0, \quad y_{\rm |j_z|max}=h_0 \pm \frac{1}{k\sqrt{5}}.
\label{eq14}
\end{equation}

\begin{figure}[h!]
	\begin{center}
		\includegraphics[width=\hsize]{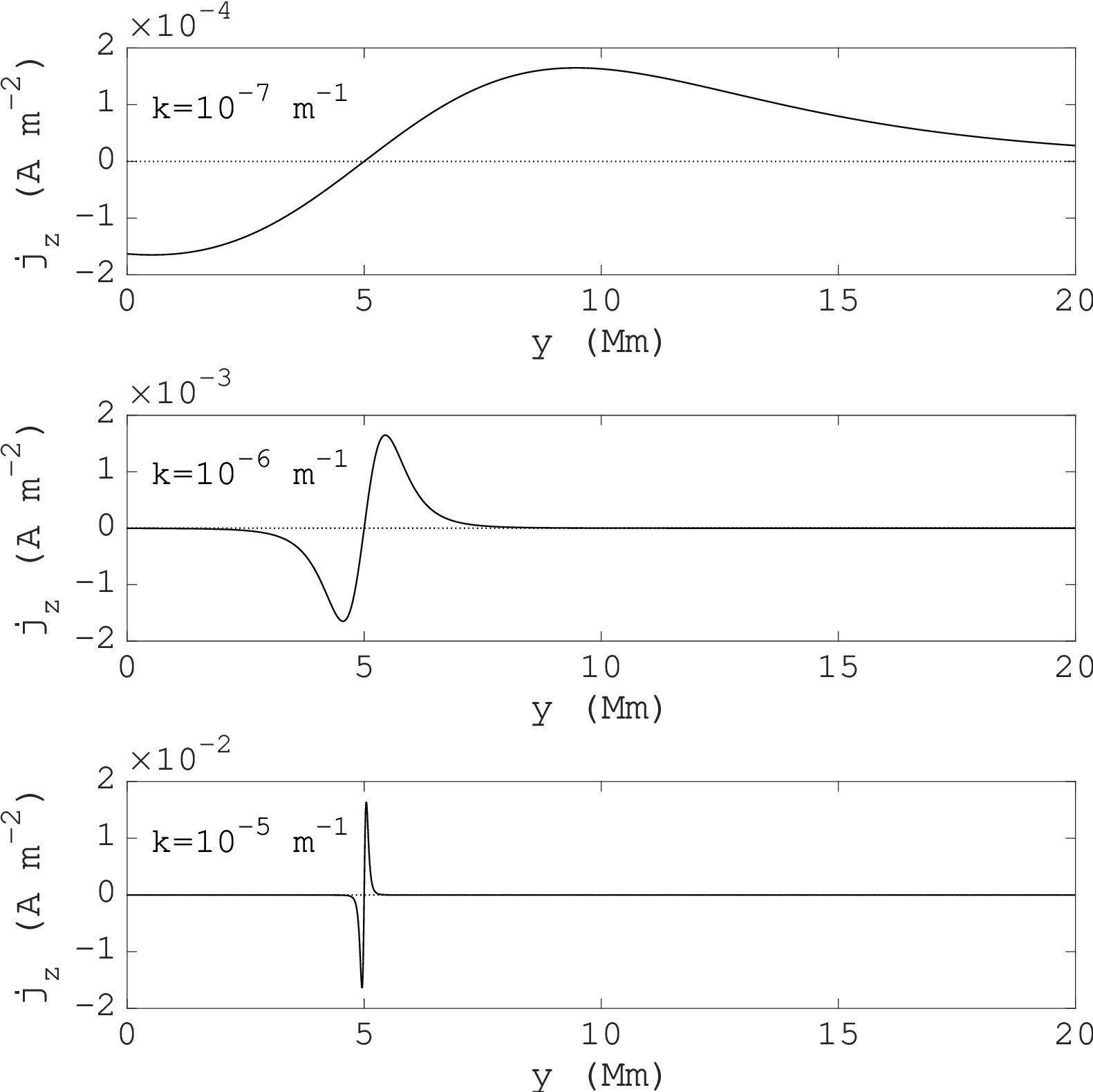}
	\end{center}
	\caption{Electric current density (z-component) along the $y$-axis $(x = 0)$ for $B_0 = 10^{-3}~\mathrm{T}$, $h_0 = 5~\mathrm{Mm}$ and three values of the parameter $k$.} \label{fig1}
\end{figure}

\begin{figure}[h!]
	\begin{center}
		\includegraphics[width=\hsize]{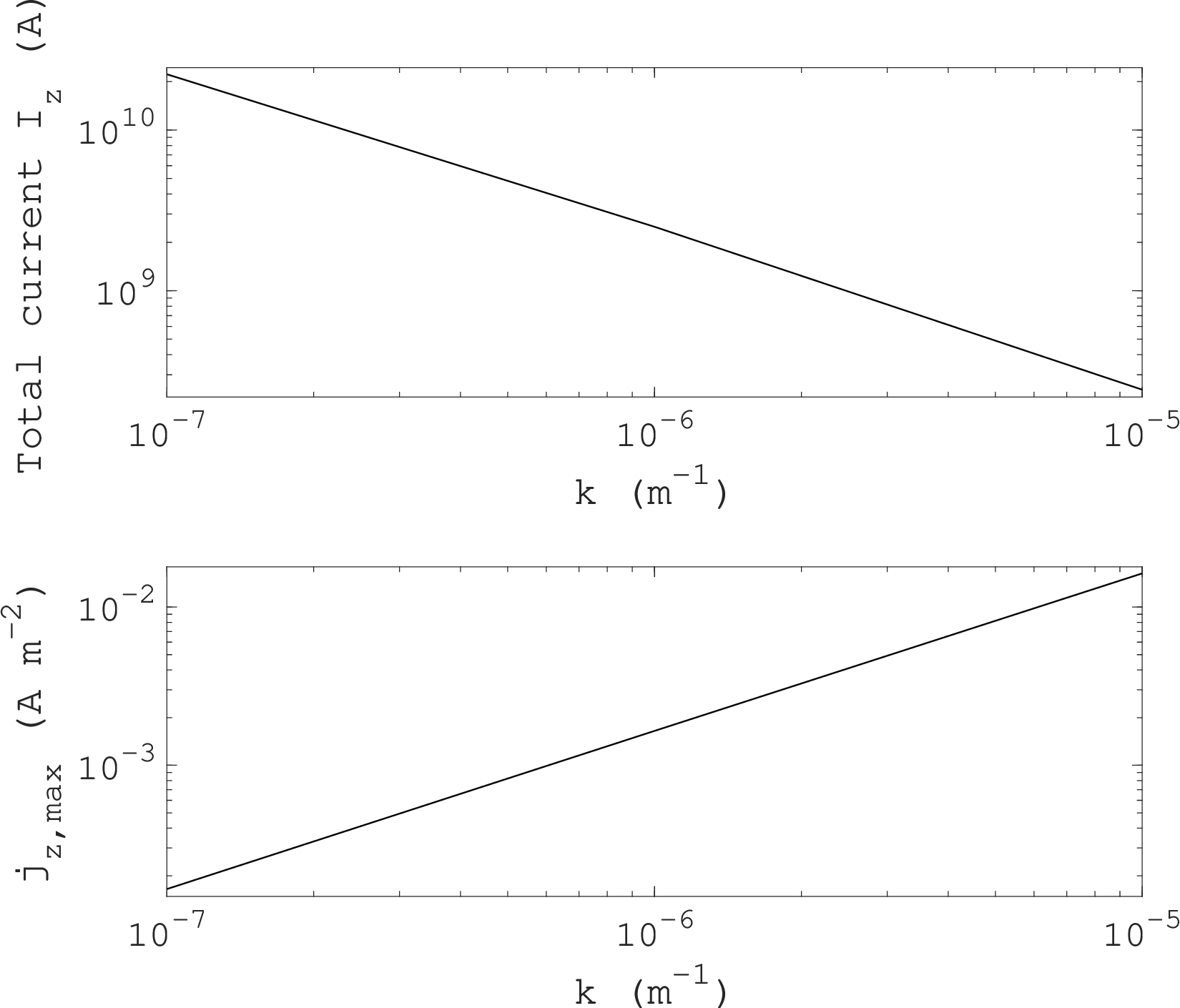}
	\end{center}
	\caption{Total electric current (upper panel) in the z-direction and maximum of the z-component of the current density (lower panel) in the filament in dependence on the parameter $k$ for $B_0 = 10^{-3}~\mathrm{T}$.} \label{fig2}
\end{figure}

For illustration, Figs.~\ref{fig1} and \ref{fig2} were created using Eq.~(\ref{eq13}). Figure~\ref{fig1} shows the $z$-component of the electric current density along the $y$-axis $(x = 0)$ for $B_0 = 10^{-3}~\mathrm{T}$, $h_0 = 5~\mathrm{Mm}$, and three values of the parameter $k$. As shown here, the maximum current density in the region above $h_0 = 5~\mathrm{Mm}$ increases with the increase of $k$, but becomes more localized. 

This trend is further confirmed in Figure~\ref{fig2}, which shows the dependence of the maximum current density on $k$. However, as $k$ increases, the total current in the filament, i.e. the current in the entire area above or below $h_0 = 5~\mathrm{Mm}$, decreases (Figure~\ref{fig2}).

\begin{figure}[h!]
	\begin{center}
		\includegraphics[width=\hsize]{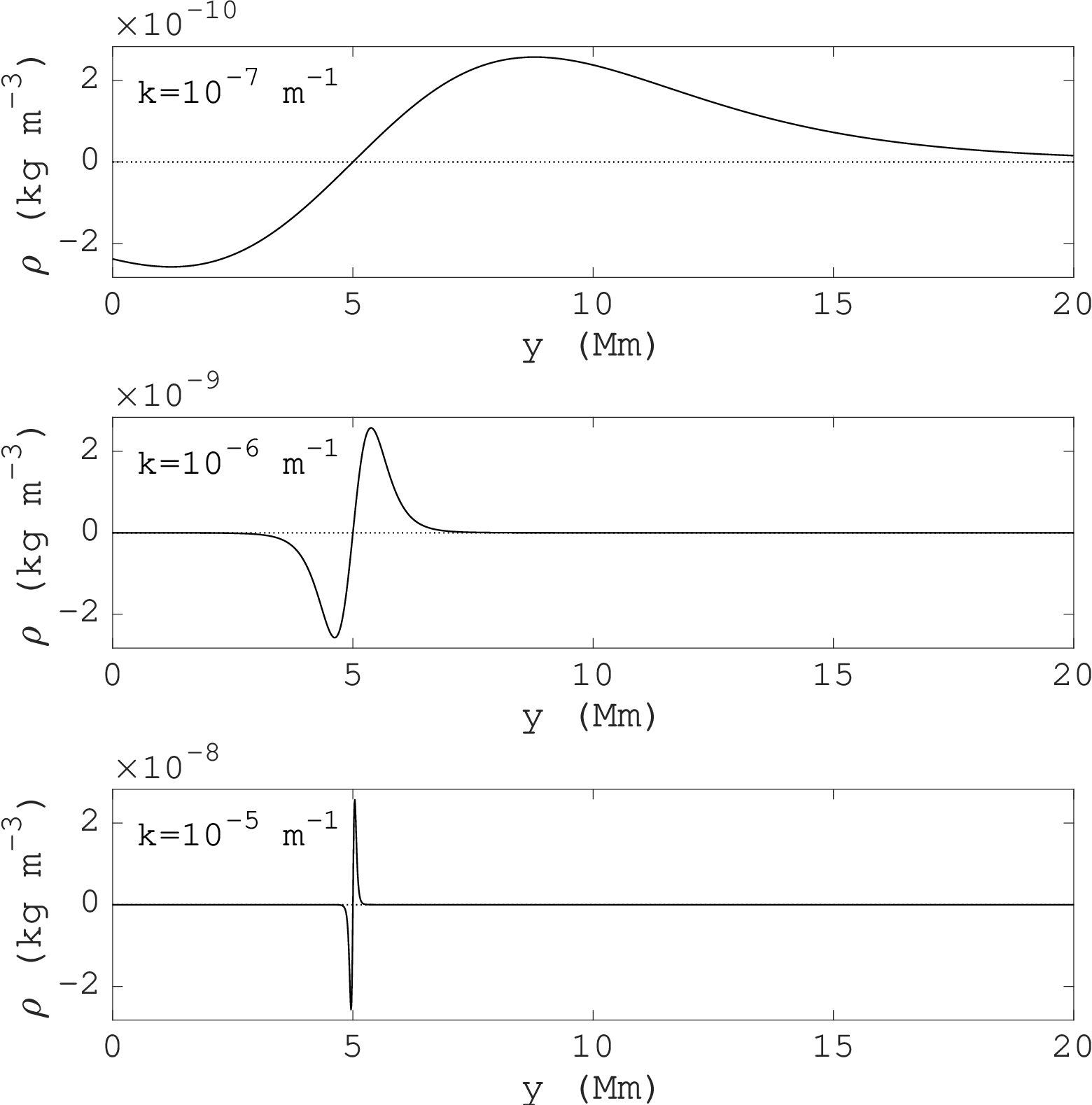}
	\end{center}
	\caption{Mass density along the $y$-axis $(x = 0)$ for $B_0 = 10^{-3}~\mathrm{T}$, $h_0~=~5~\mathrm{Mm}$ and three values of the parameter $k$.} \label{fig3}
\end{figure}

Furthermore, as shown in Figure~\ref{fig3}, the filament mass density, expressed only by the second term in the Eq.~(\ref{eq8}) and calculated along the $y$-axis $(x = 0)$, shows a similar profile as for the current density (Figure~\ref{fig1}). In this case, the distance between the maximum and minimum of the mass density is $2/(k\sqrt{7})$.

\begin{figure}[h!]
	\begin{center}
		\includegraphics[width=\hsize]{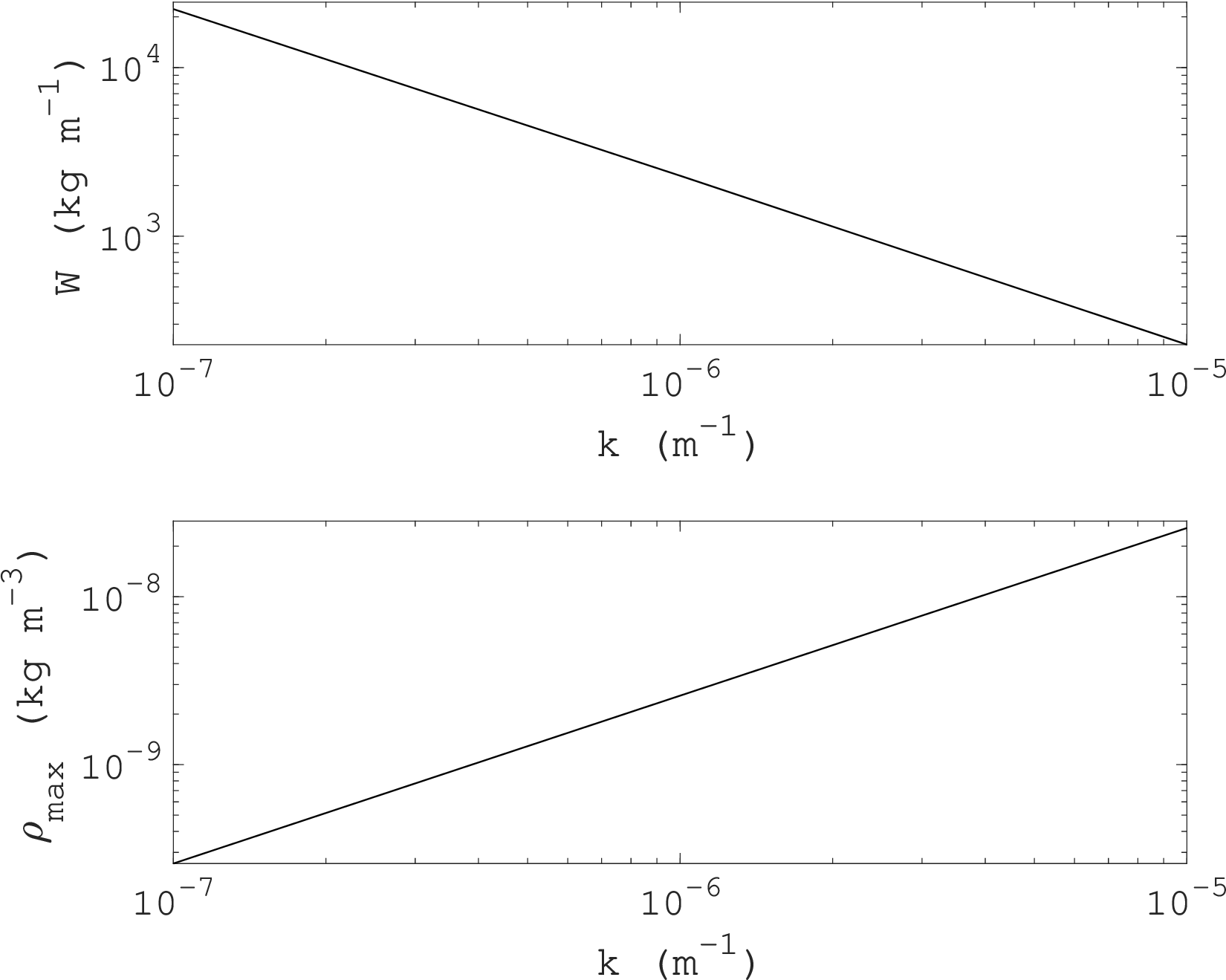}
	\end{center}
	\caption{Weight of the filament per one meter in the perpendicular direction to the $x-y$ plane (above $h_0 = 5~\mathrm{Mm}$) (upper panel) and maximum mass density (lower panel) in the filament in dependence on the parameter $k$ for $B_0 = 10^{-3}~\mathrm{T}$.} \label{fig4}
\end{figure}

Finally, Figure~\ref{fig4} shows the weight of the filament per unit length in the direction perpendicular to the $x$–$y$ plane, as well as the maximum mass density in the filament as a function of the parameter $k$ (both calculated from the second term in Eq.~(\ref{eq8}), i.e., excluding the background atmospheric density). It should be noted that the density at each single grid point, given by the sum of the second term in Eq.~(\ref{eq8}), which is negative at heights below $h_0$, and the background atmospheric density (the first term in Eq.~(\ref{eq8})), must be positive.

\section{Numerical model}
\subsection{Governing equations}
In the numerical model, we describe plasma dynamics using a set of two-dimensional, time-dependent, resistive, and compressible magnetohydrodynamic (MHD) equations formulated in the Cartesian coordinate system (see, e.g., \cite{2004prma.book.....G}). These equations account for the effects of resistivity and gravity and are expressed using the convective (Lagrangian) derivative
\begin{equation}
\frac{\mathrm{D}\varrho}{\mathrm{D}t} = -\varrho \nabla \cdot \mathbf{v}
\label{eq15}
\end{equation}
\begin{equation}
\varrho \frac{\mathrm{D}\mathbf{v}}{\mathrm{D}t} = -\nabla p + \mathbf{j} \times \mathbf{B} + \varrho \mathbf{g},
\label{eq16}
\end{equation}
\begin{equation}
\rho \frac{\mathrm{D}\varepsilon}{\mathrm{D}t} = -p \nabla \cdot \mathbf{v} + \eta |\mathbf{j}|^2,
\label{eq17}
\end{equation}
\begin{equation}
\frac{\mathrm{D}\mathbf{B}}{\mathrm{D}t} = (\mathbf{B} \cdot \nabla)\mathbf{v} - \mathbf{B} (\nabla \cdot \mathbf{v}) - \eta \nabla \times \mathbf{j},
\label{eq18}
\end{equation}
\begin{equation}
\nabla \cdot \mathbf{B} = 0.
\label{eq19}
\end{equation}
Here, $\varrho$ is the mass density, $\mathbf{v}$ is the plasma velocity, $p$ is the pressure, $\varepsilon$ is the specific internal energy, $\mathbf{B}$ is the magnetic field, $\eta$ is the electrical resistivity and $\mathbf{g}$ represents the gravitational acceleration due to the Sun.

Generally, the terms expressing the radiative losses, $R_{\rm{loss}}$, thermal conduction, $T_{\rm{cond}}$, and heating, $H$, should be added to the set of MHD equations. Because simulations are performed in an energetically closed model without any external sources, we assume that radiative losses and thermal conduction are fully compensated by the heating, $H$; in other words, $R_{\rm{loss}} + T_{\rm{cond}} + H = 0$ during the whole studied process.

 The simulation domain $x \in [-10,\,10]\,\mathrm{Mm}$, $y \in [0,\,40]\,\mathrm{Mm}$  is resolved by a $2000 \times 4000$ grid, corresponding to the smallest spatial resolution $\Delta x = \Delta y = 0.01\,\mathrm{Mm}$. This resolution sufficiently covers the studied magnetic structures. User-defined boundary conditions are applied, with the initial values of the variables fixed at the boundaries.

To visualize the simulation data, we used the VisIt software package, which is a free interactive parallel visualization and graphical analysis tool for viewing scientific data, see, e.g. \cite{HPV:VisIt}, and Matlab \cite{mathworks2024}.

\subsection{Model’s initial equilibrium state}
For the filament model in all the following simulations we set the values of the magnetic field $B_0$ and the external magnetic field $B_\mathrm{ex}=10^{-3}~\mathrm{T}$, the height of the central double structure $h_0 = 4.7~\mathrm{Mm}$, the parameter $\alpha = 0.5$, and the parameter $k~=~5~\times~10^{-7}~\mathrm{Mm}^{-1}$.

As follows from Eqs.~(\ref{eq5}) and (\ref{eq6}), the mass density and pressure consist of two components:
a) the contribution from the solar atmosphere without the filament ($\rho_h(y), p_h(y)$), and
b) the contribution from the filament itself (the remaining terms in these equations).
For the atmospheric component, we firstly used the temperature profile from the C7 (VAL-C) model of the solar atmosphere \citep{1981ApJS...45..635V,2008ApJS..175..229A}. 
However, in this case, the magnetic field lines were not sufficiently anchored during the filament ejection.

Therefore, we proposed a modified temperature profile that preserves the temperature jump characteristic of the real solar atmosphere (a ratio of about 200), that is, approximately $6000~\mathrm{K}$ in the lower layers (photosphere) and $1.2~\mathrm{MK}$ in the upper layers (corona), but with an abrupt transition in our defined "transition region".

To avoid numerical errors caused by the steep temperature gradient in this area, we expressed the temperature using a hyperbolic tangent function:
\begin{equation}
T(y) = T_{\mathrm{ph}} + \frac{1}{2}(T_{\mathrm{cor}} - T_{\mathrm{ph}}) \cdot \left[1 + \tanh\left(\frac{y - h_{\mathrm{TR}}}{w_\mathrm{TR}} \right) \right],
\label{eq20}
\end{equation}
where $T_{\mathrm{ph}} = 6000 ~\mathrm{K}$ is the photospheric temperature, $T_{\mathrm{cor}}~=~1.2~\mathrm{MK}$ is the coronal temperature, $h_{\mathrm{TR}}~=~5.0~\mathrm{Mm}$ is the height of the transition region as defined in our model and $w_\mathrm{TR}~=~0.01~\mathrm{Mm}$ is the width of this model-defined transition region.

\begin{figure}[h!]
	\begin{center}
		\includegraphics[width=\hsize]{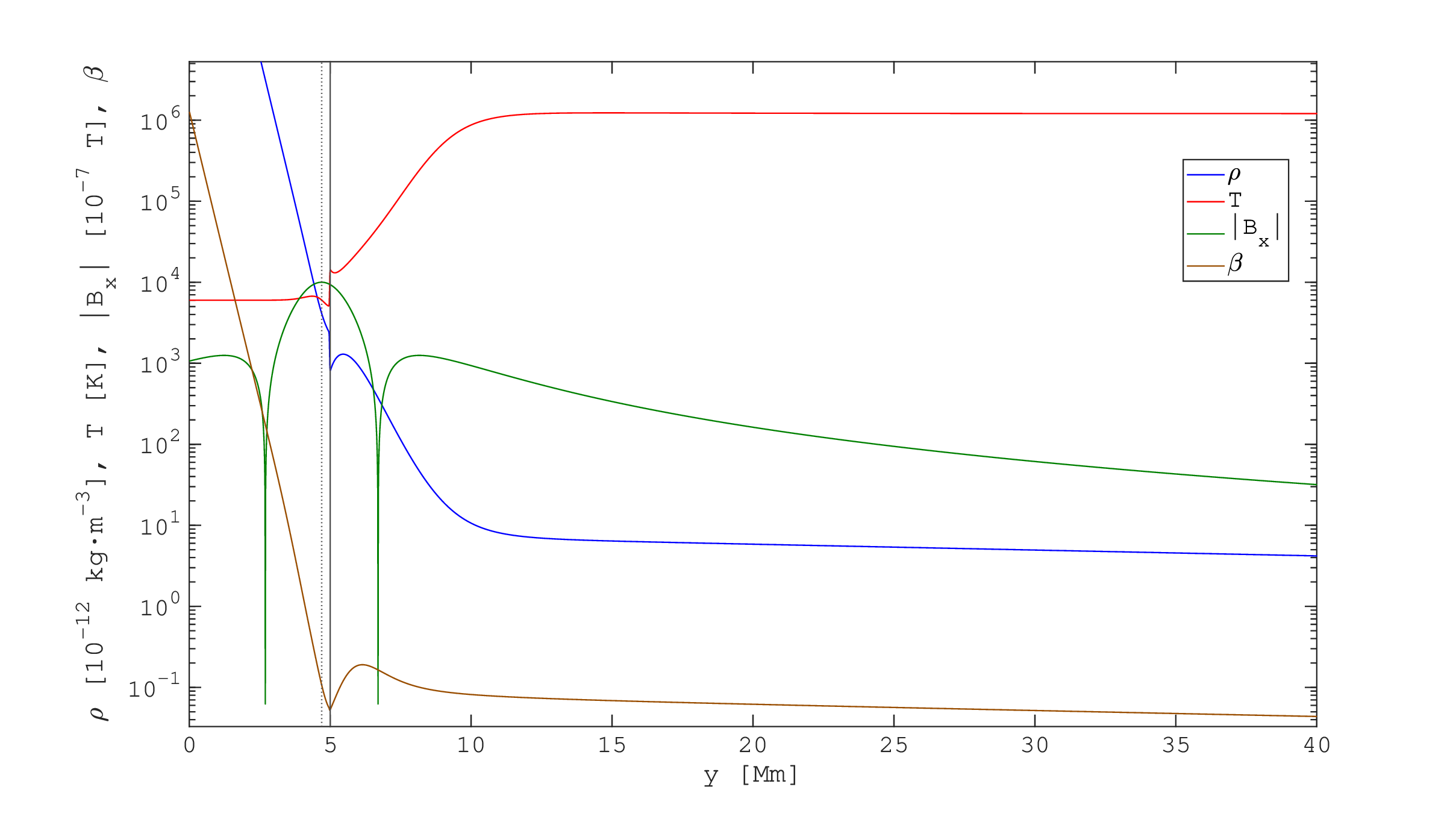}
	\end{center}
	\caption{Vertical profiles of the absolute value of $B_x$ component of magnetic field (green line), mass density $\rho$ (blue line), temperature T (red line), and plasma beta parameter $\beta$ (brown line) along the axis of symmetry $(x = 0~\mathrm{Mm})$ in the initial equilibrium state for $B_0$ = 10$^{-3}$ T and $k = 5\times10^{-7}~\mathrm{m^{-1}}$. The solid black vertical line shows the location of the transition region, $h_{\mathrm{TR}} = 5~\mathrm{Mm}$, as defined in our model, while the dotted vertical line denotes the midpoint between the filaments $h_0 = 4.7~\mathrm{Mm}$. Note that the $y$-axis is on a logarithmic scale.} \label{slice_0000}
\end{figure}

In Figure~\ref{slice_0000}, we present the initial equilibrium state ($t=0~\mathrm{s}$) for four quantities from our numerical model: the density (blue), the temperature (red), the absolute value of the magnetic field component $B_x$ (green line), and the plasma beta parameter $\beta$ (brown line). The figure also shows two vertical lines indicating the position of the transition region $h_\mathrm{TR}$, as defined in our model (solid black line), and the location of the center of the double-structure of the filament $h_0$ (dashed line). The mass density decreases from the lower boundary of the simulation box in agreement with the hydrostatic decrease in the solar gravitational field. At the location of $h_\mathrm{TR}$, where the temperature jump occurs, the mass density begins to increase, forming a bump (modeled filament) with a maximum density of roughly $1.3 \times 10^{-9}~\mathrm{kg\,m^{-3}}$, and then decreases again to a value of about $1 \times 10^{-12}~\mathrm{kg\,m^{-3}}$, typical value in the solar corona. The model temperature follows Eq.~(\ref{eq20}). Above $h_\mathrm{TR}$, in the bump region, which represents the modeled filament, the temperature drops to approximately $13000~\mathrm{K}$ and then increases to approximately $T_{\mathrm{cor}} = 1.2~\mathrm{MK}$, typical of the solar corona.

\section{Results}
In this section, we present the results of numerical simulations aimed at exploring the ejection of the filament owing to its destabilization. Prior to applying any perturbations, we performed a baseline simulation using the equilibrium configuration to verify numerical stability and ensure that the magnetic field and plasma
remain stationary in the absence of external changes.

We then considered two destabilizing scenarios: (a) an increase of the electric current in the filament, and (b) a decrease of the filament mass density. In our model, these changes are imposed suddenly for simplicity. In reality, solar processes are more gradual. However, if the current increase or density reduction occurs gradually while the entire filament system is confined by an overlying magnetic field that then suddenly opens, the resulting ejections could resemble to the following simulations.

\begin{figure*}
  \includegraphics[width=\hsize]{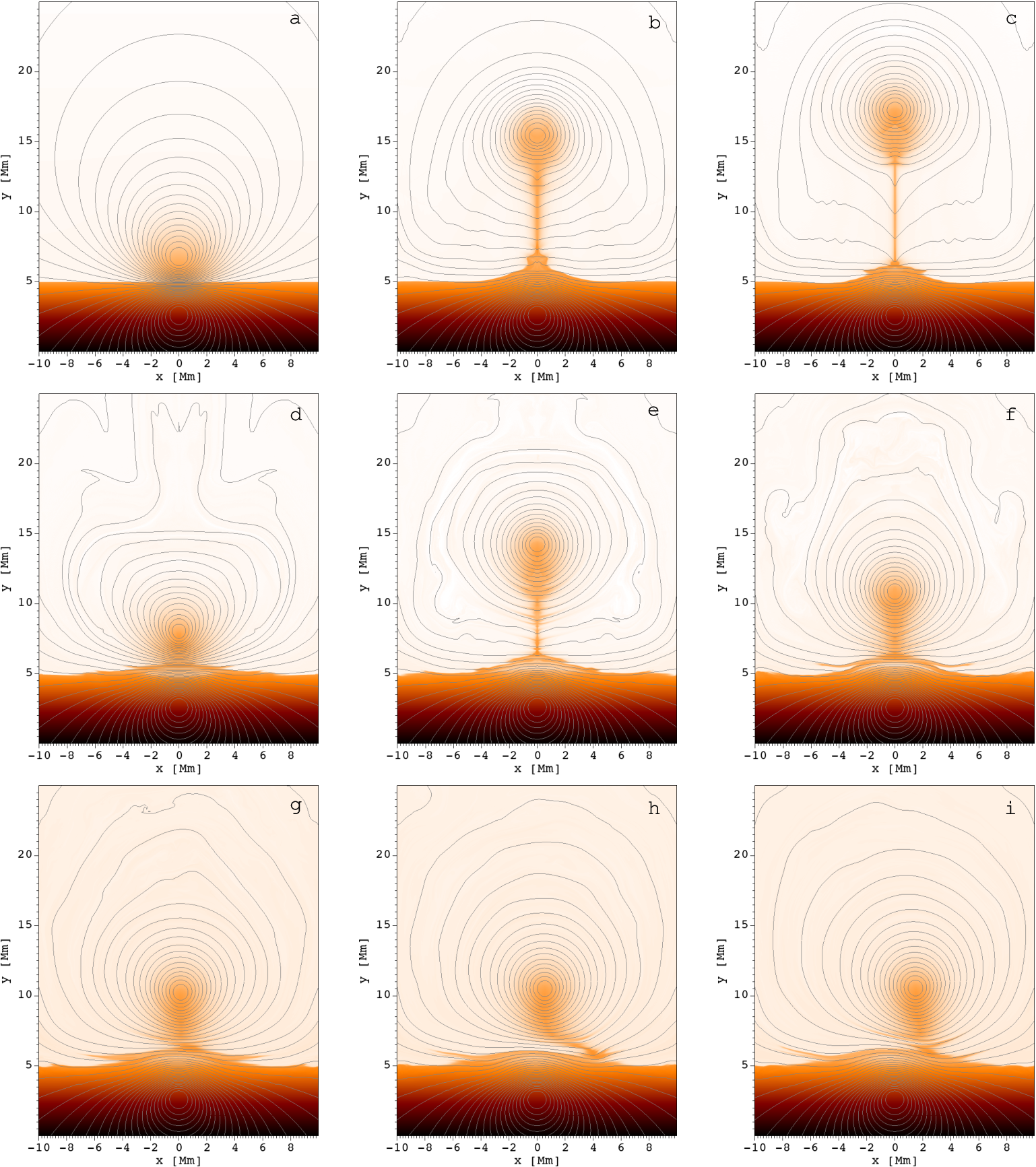}
     \caption{Time evolution of mass density and magnetic field lines, using the vector potential, represented by gray solid lines, for nine different times $t = 0, 168, 280, 628, 897, 1121, 2240, 2800 $ and $3360~\mathrm{s}$ (a - i, respectively) - the case with the increased electric current.}
     \label{evol_4B0}
\end{figure*}

\begin{figure*}
\begin{center}
    \includegraphics[width=\hsize]{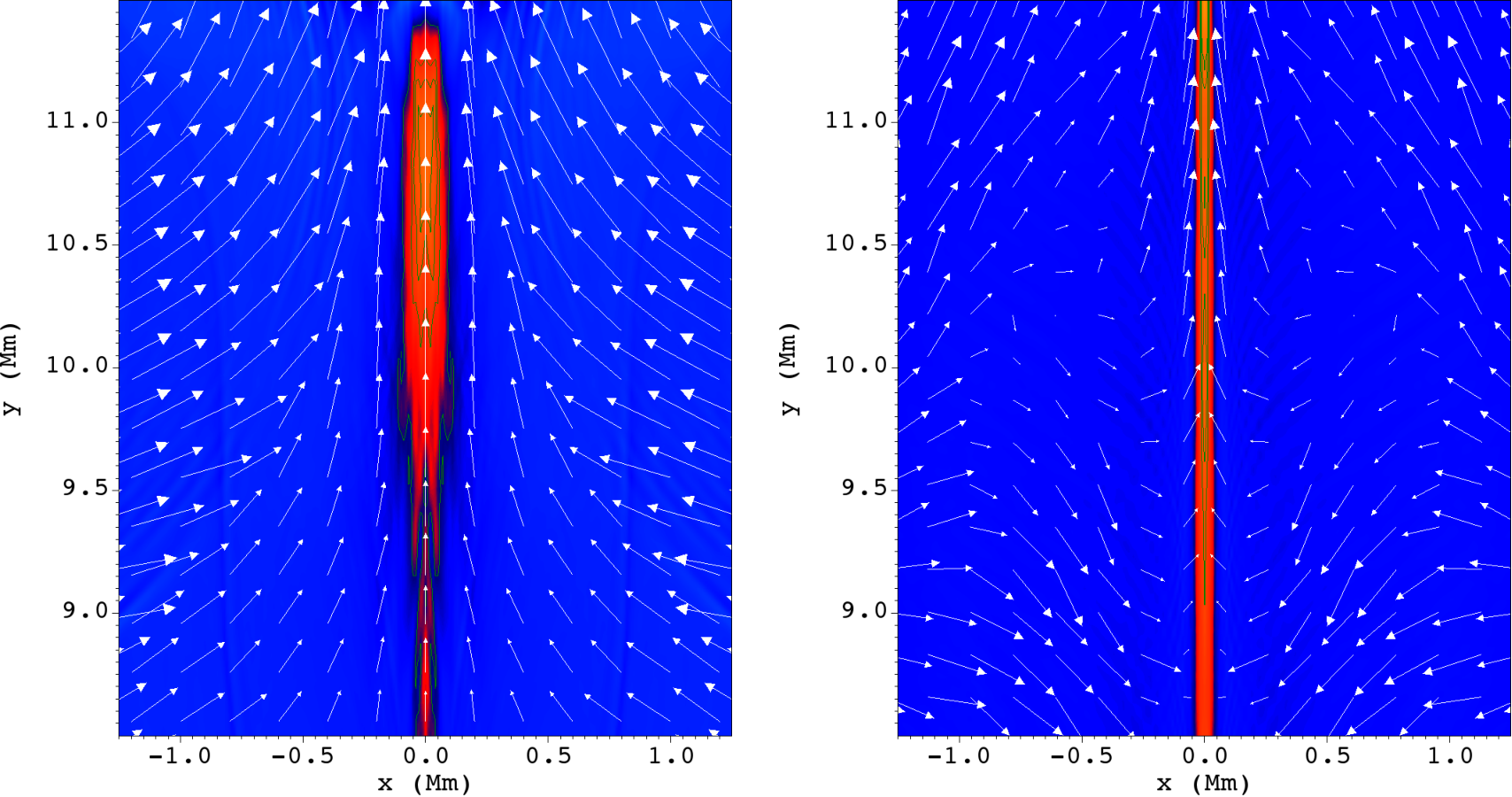}
    \caption{Plasma flows (white arrows) around the current sheet, expressed here through the current density $j_z$ at two times $t$ = 168 s (left) and 280 s (right) in the case with $4\,B_0$. These panels present details at the same times as the panels b) and c) in Figure~\ref{evol_4B0}.} \label{plasma_flows}
\end{center}
\end{figure*}

\subsection{Failed ejection and oscillations of the filament induced by increasing the electric current}
In this case, we start with the model in equilibrium as described above. We then increase both the upper and lower electric currents and analyze the resulting dynamical evolution of the system.

The increased current in the model was achieved by increasing $B$ compared to its initial value $B_0$. Other parameters were not changed. As an example, here we present the case where the magnetic field strength $B_0$ is increased by a factor of $4$.

The results of the simulations are presented in Figure~\ref{evol_4B0}. The figure shows the temporal evolution of the mass density (brown colour scale) and the magnetic field, which is represented by using the magnetic vector potential and is illustrated in the figure with solid grey lines.

The first panel (a) shows the distribution of these quantities in the initial state. The destabilization of the system leads to the ejection of the filament upward, caused by the increased magnetic force acting between the upper and lower currents. The lower currents remain stationary as they are sufficiently anchored in the dense layers of the atmosphere. Panel~(b) shows the situation at time $t=168~\mathrm{s}$, when the filament is ejected and rises upward. Panel~(c) corresponds to the state when the filament has already reached its maximum height, at time $t=280~\mathrm{s}$.

Then, the filament starts to move downward toward the dense layers of the solar atmosphere. This is illustrated in panel (d), corresponding to time $t=628~\mathrm{s}$, when the filament approaches the moment of reaching these dense atmospheric layers. Then it remains nearly stationary for a short time before starting to move upward again. This is seen in panel (e) at $t=897~\mathrm{s}$, when the filament reaches a new maximum height, about $3~\mathrm{Mm}$ lower than during the initial ejection. Panel (f) at $t=1121~\mathrm{s}$ shows the filament again descending toward dense atmospheric layers. In the following times, this cycle repeats with the period of about $600~\mathrm{s}$. However, as seen in panel (g) at $t=2240~\mathrm{s}$ the density structure beneath the filament changed from the density distribution symmetric along the vertical axis ($x=0$) to that with some asymmetry. 
This asymmetry in subsequent times leads to a displacement of the filament off the vertical axis ($x=0$) (panels (h) and (i) at $t=2800~\mathrm{s}$ and $t=3360~\mathrm{s}$).
The evolution of the filament ejection and its fall down is shown in the movie, enclosed in the paper. To verify that the size of the numerical domain does not influence the magnetic field configuration and the maximum height reached by the filament, we performed an additional comparative simulation. The numerical box was extended by a factor of two in the vertical ($y$) direction, while all other parameters were kept unchanged. The maximum filament height was then compared with that obtained in the reference case shown in Figure~\ref{evol_4B0}. No significant difference was found, confirming that the original vertical extent of the numerical domain is sufficient and does not affect the results.

Below the ejected filament, a current sheet is formed. To examine this process in more detail, Figure~\ref{plasma_flows} shows plasma flows (vectors) around the current sheet, represented by the current density $j_z$, at two times: $t = 168$ s (left) and $t = 280$ s (right) for the case with $4\,B_0$. These can be compared with panels b) and c) of Figure~\ref{evol_4B0}. The panels in Figure~\ref{plasma_flows} illustrate how plasma from both sides of the current sheet is drawn into its region and directed upward. The structured current sheet (with magnetic islands) and the associated plasma velocity pattern, most evident in the right panel), indicate magnetic reconnection. This reconnection disrupts the magnetic field lines connecting upper and lower currents.

\begin{figure}[h!]
	\begin{center}
		\includegraphics[width=\hsize]{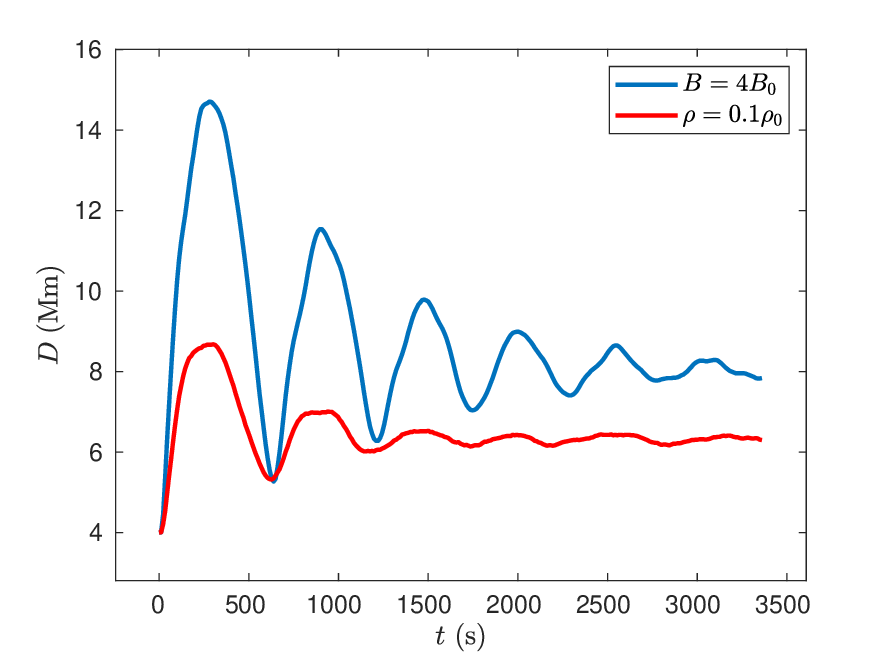}
		\end{center}
	\caption{Time evolution of the distance $D$ between the upper and lower magnetic-structure centers (with $B_x=B_y=0$) for two cases with $4\,B_0$ (blue line) and $0.1\,\rho_0$ (red line).} \label{h-time}
\end{figure}

\begin{figure}[h!]
	\begin{center}
        \includegraphics[width=\hsize]{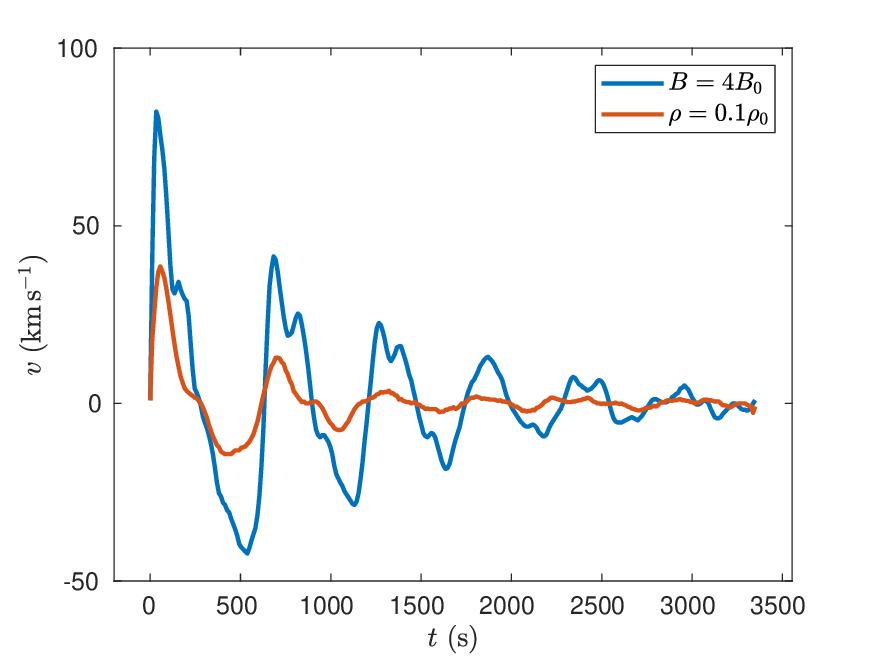}
		\end{center}
	\caption{Temporal evolution of the filament ejection velocity $v$ during its ejection and subsequent oscillations for the case with $4\,B_0$ (blue) and the case with $0.1\,\rho_0$ (red).} \label{v-time}
\end{figure}

The process of filament ejection and its oscillation is also illustrated in Figure~\ref{h-time}, which shows the temporal evolution of the distance $D$ between the center of the upper and lower magnetic structures (with $B_x=B_y=0$). In this figure, two curves are shown, with the blue curve corresponding to the case of increased electric current. The period of this oscillation is approximately $600\,\mathrm{s}$. The corresponding velocity of the filament ejection and the following oscillations in both cases are shown in Figure~\ref{v-time}. As shown here, the maximum velocity of filament ejection was $80~\mathrm{km\,s}^{-1}$ in the current-increase case.

\begin{figure}[h!]
	\begin{center}
  	\includegraphics[width=\hsize]{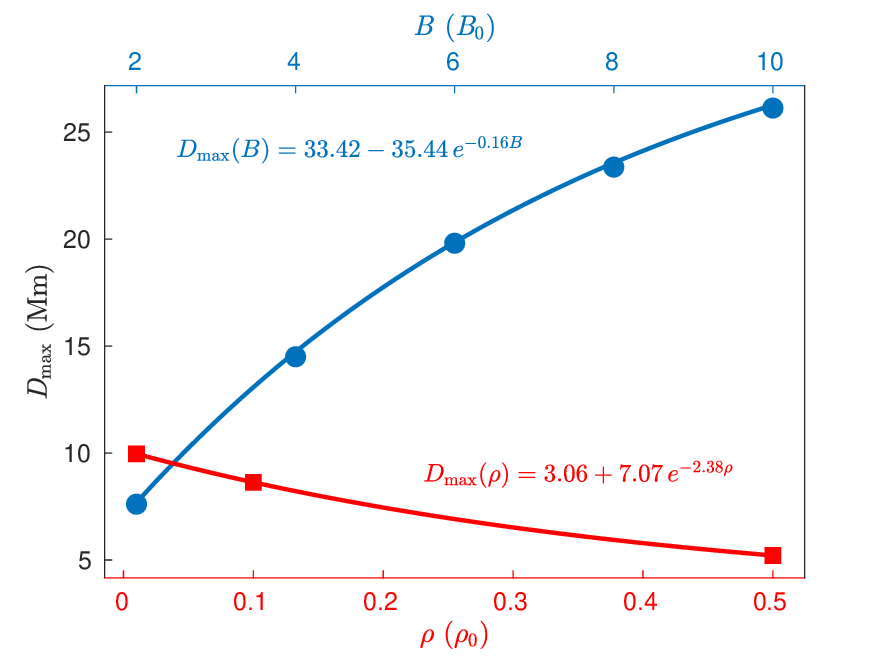}
	\end{center}
    \caption{Maximum distance $D_\mathrm{max}$ between the upper and lower magnetic-structure centers (with $B_x=B_y=0$) as a function of multiplication of the initial magnetic field  ($2\,B_0$, $4\,B_0$, $6\,B_0$, $8\,B_0$, and $10\,B_0$, with $B_0=10^{-3}$~T; top $x$-axis) and as a function of the filament density dilution ($0.01\,\rho_0$, $0.1\,\rho_0$, and $0.5\,\rho_0$, where $\rho_0$ is the initial density in the filament; bottom $x$-axis).}
    \label{h-xB0}
\end{figure}

In Figure~\ref{h-xB0}, the blue points fitted by the exponential function again correspond to the case of increased electric current. In this figure, we can see the dependence of the maximum distance $D_\mathrm{max}$ reached between the centers of the upper and lower magnetic structures (with $B_x=B_y=0$). The cases correspond to multiples of the magnetic field $2\,B_0$, $4\,B_0$, $6\,B_0$, $8\,B_0$, and $10\,B_0$ (where $B_0 = 10^{-3}~\mathrm{T}$).

\begin{figure*}[h!]
\begin{center}
  \includegraphics[width=18.5cm]{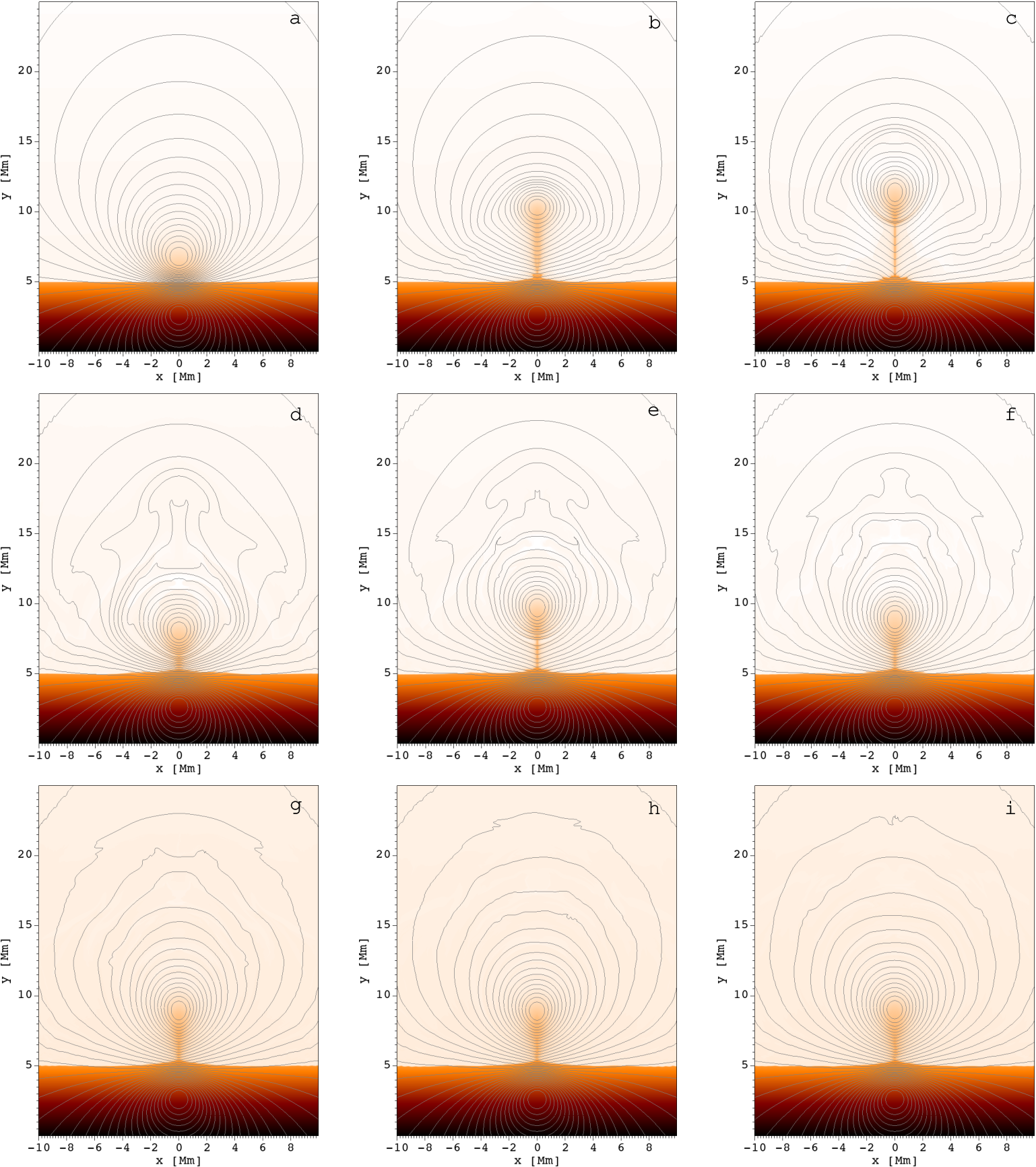}
\end{center}
     \caption{Time evolution of mass density and magnetic field lines, using the vector potential, represented by black solid lines, for nine different times $t = 0, 123, 280, 628, 897, 1121, 2240, 2800$ and $3360~\mathrm{s}$ (a - i, respectively) - the case with the evacuation of filament to $10~\%$ of initial value of the mass density.}
     \label{evol_01rho0}
\end{figure*}

\subsection{Failed ejection and oscillations of the filament induced by decreasing the mass density}
In the second scenario, we again start from the equilibrium model but artificially reduce the mass density of the filament and numerically track its subsequent evolution.

In a real filament, such a density decrease may result from plasma heating in a localized region. Heating increases the pressure and decreases the density, which reduces the local weight of the filament and disrupts its equilibrium. The affected segment then begins to rise, while plasma drains from it under the influence of gravity. This further decreases the density in the region and enhances the upward motion.

In Figure~\ref{evol_01rho0}, we again show the temporal evolution of the density and magnetic field at nine different times, similar to the first studied case described above. Calculations were performed for three different values of the reduced mass density in the filament, namely $0.01\,\rho_0$, $0.1\,\rho_0$, and $0.5\,\rho_0$. The results presented here correspond to the case with $\rho = 0.1\,\rho_0$.

In panel (a), corresponding to time $t=0~\mathrm{s}$, we again see the initial state. In the following panel (b), at $t=123~\mathrm{s}$, the filament begins to rise until it reaches the time $t=280~\mathrm{s}$, as shown in panel (c). Afterwards, the upper filament starts to move downward toward the lower layers of the atmosphere, which it reaches at approximately $t=628~\mathrm{s}$, as shown in panel (d). Then it attains another maximum distance between the centres of the two magnetic structures at $t=897~\mathrm{s}$, as illustrated in panel (e). Panel (f) at $t=1121~\mathrm{s}$ shows the state when the filament again approaches the lower atmospheric layers. This oscillation process repeats with a period of about $600~\mathrm{s}$. In the following times, contrary to the case of $4\,B_0$, the filament oscillates up to a stable state along the vertical axis ($x=0$) (panels (g), (h) and (i) in $t=2240~\mathrm{s}$,  $t=2800~\mathrm{s}$ and $t=3360~\mathrm{s}$.  
The evolution of filament ejection and its subsequent oscillation are illustrated in the movie included in the paper. As in the previous case, we also performed a comparative simulation to assess the influence of the numerical domain size on the magnetic field configuration and the filament dynamics. No differences in the filament evolution or in the maximum height reached were detected. This is consistent with the fact that, in this density-reduction scenario, the filament reaches a lower height than in the current-increase case, further confirming that the original size of the numerical domain is sufficient and does not affect the results. 

In Figure~\ref{h-time}, the red curve corresponds to the case with a reduced filament density. The first feature we can observe, when compared with the case of increased electric current, is that the filament does not reach such a large height as in the current-increase scenario. The second difference is that, in the case of reduced density, the filament reaches its maximum height earlier and remains in this position for a longer duration, approximately $200~\mathrm{s}$. In contrast, in the case of increased electric current, the filament reaches its maximum height later and remains close to this maximum for only about half as long, roughly $100~\mathrm{s}$. However, the figure also shows that, despite these differences, the oscillation period of the filament is approximately the same in both cases. The velocity of filament ejection and the following oscillations in this case are shown in Figure~\ref{v-time} by the red line. In this density-reduction case, the maximum velocity of filament ejection was $40~\mathrm{km\,s}^{-1}$.

Similarly to the previous case, in Figure~\ref{h-xB0} we plot the maximum distance $D_\mathrm{max}$ reached between the centers of the two magnetic structures (red points fitted by the exponential function), but this time for three different values of the reduced initial filament density. The three values considered are $0.01\,\rho_{0}$, $0.1\,\rho_{0}$, and $0.5\,\rho_{0}$. From the plotted values, it is evident that a smaller ratio $\rho/\rho_{0}$ results in the filament reaching a higher height.

\section{Discussion}

\begin{figure}
	\begin{center}
        \includegraphics[width=\hsize]{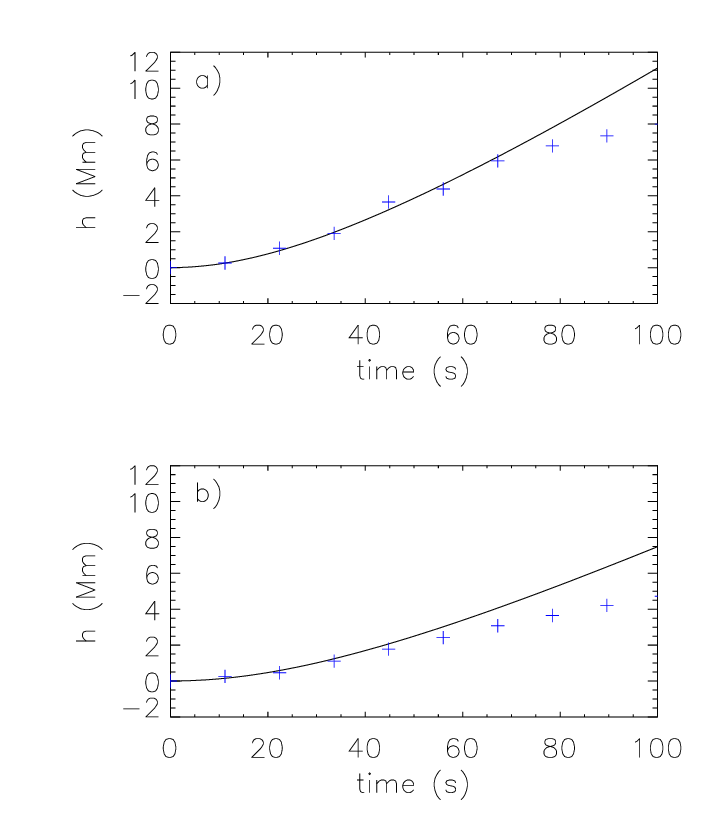}
	\end{center}
	\caption{Time evolution of the height $h$ of the upper current in the ideal vacuum model (solid lines) and of the filament’s maximum current density in the MHD model (crosses), both measured relative to their initial equilibrium positions, over the $0-100~\mathrm{s}$ interval, for the two cases with $4\,B_0$ (panel a) and $0.1\,\rho_0$ (panel b).}
	\label{comp}
\end{figure}

To understand the origin of the found failed ejection and oscillations, we consider a simple analytical model consisting of two parallel, oppositely directed currents interacting in a vacuum. In equilibrium, the repulsive magnetic force between the currents is balanced by the weight of the upper current (the filament). The lower current is fixed in position. The system is destabilized either by increasing the currents or by reducing the mass of the filament. A detailed description of this model is provided in the Appendix.

As shown in the Appendix, the oscillations arise from the interplay between two forces: the Lorentz force and the filament's weight. A rough comparison between the oscillations of the distance $D$ obtained from the MHD simulations and the oscillations of the height $h$ of the upper current in the ideal vacuum model shows that both the amplitudes and the periods differ significantly (see Figures \ref{graph1}, \ref{graph2}, and \ref{h-time}).
	
Therefore, let us compare the evolution of the height of the upper current above its initial position in the ideal vacuum model with the height of the filament’s maximum current density above its initial position in the MHD model at the very beginning of the ejection.
The comparison presented in Figure \ref{comp}a shows that the evolution of the height in both models destabilized by a current increase is similar up to about $65~\mathrm{s}$. After this time, however, the evolutions begin to diverge: the height in the MHD model becomes progressively smaller than in the ideal vacuum model. A similar trend is seen in Figure \ref{comp}b, where the heights in both models are compared for the case with reduced filament density. In this case, the divergence appears to begin around $50~\mathrm{s}$.     

This raises the question of what causes these differences in evolution. The answer lies in plasma processes that are absent in the vacuum model, as well as in the different spatial distributions of currents. In the ideal vacuum model, the currents are localized at specific points, whereas in the MHD model, they are concentrated in some regions. Note also that, in the MHD model, a large region surrounding the maximum current density moves during oscillations as a coherent structure.
	
In the MHD model, magnetic reconnection is forced by the interaction between the upper and lower currents. This reconnection disrupts some of the magnetic field lines that link the filament current with its sub-photospheric counterpart, thereby reducing the repulsive force between them.
Magnetic reconnection begins at the onset of the ejection, after which a current sheet forms beneath the filament. At this stage, the reconnection appears to become more effective, as indicated by the divergence of the filament's current trajectories in the MHD and ideal vacuum models.

Additionally, in the case where destabilization is caused by a reduction in filament mass, the plasma beta parameter is close to unity ($0.33$ at a height of $6~\mathrm{Mm}$; see Figure \ref{slice_0000}). Under such conditions, motions driven by the Lorentz force are suppressed \citep{2000mare.book.....P}. Moreover, comparing the $0.1\,\rho_0$ and $4\,B_0$ cases, the parameter $\chi$ is less than $\xi^2$. These facts may explain why the divergence of the filament's current trajectories begins earlier and why the maximum height reached by the filament is smaller than in the case of destabilization due to an increased current. It should be noted that in the case destabilized by a fourfold increase in current, the plasma beta parameter is $16$ times lower than in the mass-reduction case; consequently, this effect is much less significant.

Comparing the time evolution of the filament in the $4\,B_0$ and $0.1\,\rho_0$ cases in Figures \ref{evol_4B0} and \ref{evol_01rho0}, we see that the evolution in the $4\,B_0$ case is more dynamic. In the $0.1\,\rho_0$ case, the filament settles into a stable state after its damped oscillations, whereas in the $4\,B_0$ case the filament evolves into a state that is displaced from the vertical axis ($x=0$). This appears to be caused by some instability of the whole system in the $4\,B_0$ case.	

Maximum velocities in the ideal-vacuum model and the MHD models also differ significantly (Figure~\ref{v-time}). The velocity profiles vary as well. In the ideal-vacuum model, the velocity profile is smooth and exhibits a single maximum per period, whereas in the MHD model several secondary maxima appear during the filament’s descent. We suggest that these velocity variations arise because some regions beneath the filament become compressed as it falls.

\section{Conclusions}

In this work, first, we studied the Solov'ev model \citep{2010ARep...54...86S, 2012Ge&Ae..52.1062S} analytically. We derived relations for the electric current density and plasma density and showed their spatial distributions as functions of the model parameters.

Then we investigated the stability and ejection of a gravity-balanced, current-carrying filament using 2D MHD simulations with the \texttt{Lare2D} code. The initial configuration was based on the Solov'ev model, which provides a useful framework for analyzing the interplay between magnetic and gravitational forces in filament equilibrium. To study processes after filament destabilization, we considered two types of perturbation. In the first case, the electric currents in the filament system were increased, enhancing the repulsive Lorentz force and driving the filament upward from its initial equilibrium. In the second case, the plasma density within the filament was reduced relative to its equilibrium value, thereby weakening the stabilizing influence of gravity and allowing the filament to rise 

In both cases of destabilization, the filament was ejected upward. After reaching a certain altitude, it stopped and then began to move downward, followed by a return upward, thus performing damped oscillations with a period of about $600\,\mathrm{s}$. The maximum velocity of filament in the first ejection phase reached about $80~\mathrm{km\,s}^{-1}$ in the current-increase case and about $40~\mathrm{km\,s}^{-1}$ in the density-reduction case. The maximum altitude attained by the filament during ejection increases with either higher electric current or lower filament density. These dependencies can be described by exponential functions.

As seen in Figure~\ref{h-time}, the filament during oscillations converges to a position that is at a higher height than its initial location. This is because the magnetic reconnection accumulates plasma beneath the filament, thereby supporting it against gravity at a higher altitude.
        
We found that the current sheet beneath the filament became more pronounced when the filament’s magnetic field lines were more effectively anchored in the dense atmospheric layer below, that is, when the filament structure was embedded deeper in this dense region. Specifically, when the filament structure is embedded more deeply, the magnetic field lines on opposite sides of the current sheet are more antiparallel.

In our MHD model, the current increase or density reduction that destabilizes the filament is suddenly introduced for the sake of simplicity. In reality, solar processes can be more gradual. However, if the current increase or density reduction occurs gradually while the entire filament system is confined by an overlying magnetic field, which then suddenly opens, the resulting ejections could resemble those obtained in our simulations.

Failed eruptions \citep{2007SoPh..245..287G,2024SoPh..299...81M}, which begin as typical eruptions but, for various reasons, halt in the corona after their initial acceleration, may be explained by the simulations presented in this study. The modelled behaviour (an initial rise followed by deceleration and subsequent fallback and oscillations) closely resembles the observed dynamics of such failed eruptions. For example, \citet{2022ApJ...932L...9K} reported vertical oscillations of a magnetic flux rope that appear to be similar to those obtained in our simulations, suggesting that the mechanisms described here may have direct observational counterparts.

In summary, our results demonstrate that both current enhancement and density reduction represent viable pathways leading to a failed filament ejection followed by damped oscillatory motion. The simulations reveal that magnetic reconnection beneath the filament and plasma accumulation play key roles in halting the ejection and stabilizing the system. By linking these processes to different modes of destabilization, our study provides new insight into how variations in current and plasma density can give rise to the observed diversity of filament eruption behaviours.

Overall, these findings contribute to a more comprehensive understanding of solar filament dynamics, particularly the transition between successful and failed eruptions, and highlight the importance of coupling between magnetic and plasma processes in shaping the evolution of solar eruptive events.

\begin{acknowledgements}
The work in this article is part of the project DynaSun, that has received funding under the Horizon Europe programme of the European Union under grant agreement (no. 101131534). Views and opinions expressed are, however those of the author(s) only and do not necessarily reflect those of the European Union and therefore, the European Union cannot be held responsible for them. MK acknowledges institutional support from the project RVO-67985815. This work was supported by the Ministry of Education, Youth and Sports of the Czech Republic through the e-INFRA CZ (ID:90254).
\end{acknowledgements}

\bibliographystyle{aa}
\bibliography{ref}

\begin{appendix}
	\section{Ideal vacuum model}
	Let us consider the ideal case of two parallel currents $I$, oriented in the opposite direction and separated by a distance 
$d$ in vacuum. The lower current is fixed in position, while the upper current (a filament) has a mass per unit length 
$m$. In equilibrium, the repulsive magnetic force between the currents balances the weight of the upper filament ($mg$), i.e.,
 \begin{equation}
\frac{\mu_0 I^2}{2 \pi d}	=  m g = \frac{A}{d},
\end{equation}
where A = $\frac{\mu_0 I^2}{2 \pi}$.

\begin{figure}
	\begin{center}
		\includegraphics[width=\hsize]{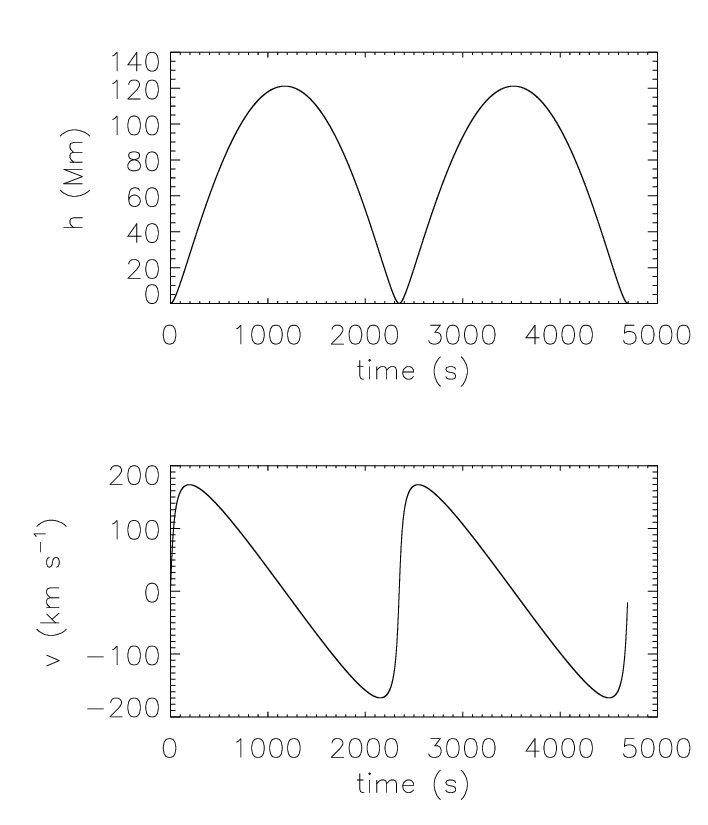}
	\end{center}
	\caption{Time evolution of the height $h$ of the upper current above its initial equilibrium position, and of its velocity $\rm v$, for the parameter $\xi = 4$ and $d=1.79~\mathrm{Mm}$, which corresponds to the $4\,B_0$ MHD case.}
	\label{graph1}
\end{figure}

\begin{figure}
	\begin{center}
		\includegraphics[width=\hsize]{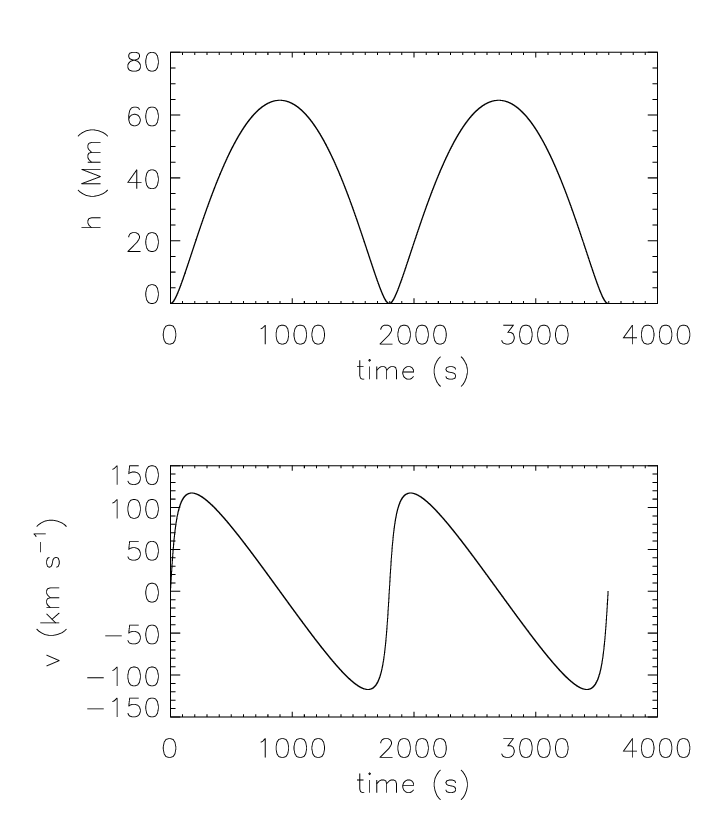}
	\end{center}
	\caption{Time evolution of the height $h$ of the upper current above its initial equilibrium position, and of its velocity $\rm v$, for the parameter $\chi = 10$ and $d=1.79~\mathrm{Mm}$, which corresponds to the $0.1\,\rho_0$ MHD case.}
	\label{graph2}
\end{figure}

Now, let us increase both currents by a factor of 
$\xi$. The weight of the upper filament remains the same. The net force acting on the filament can then be written as
\begin{equation}
F = \frac{\xi^2 A}{d + h} - \frac{A}{d},
\end{equation}
where $h$ denotes the vertical displacement (height) of the upper filament from its equilibrium position.

Hence, the equation of motion for the upper filament becomes
\begin{equation}
	m \frac{\mathrm{d}^2 h (t)}{\mathrm{d} t^2 } = \frac{\xi^2 A}{d+h} -\frac{A}{d}.
\end{equation}
Substituting  $m =\frac{A}{dg}$ simplifies the equation to	
\begin{equation}
	\frac{\mathrm{d}^2 h (t)}{\mathrm{d} t^2 } = \frac{\xi^2 d g}{d+h} - g.
	\label{a4}
\end{equation}

The right-hand side of this equation depends only on 
$h$. Multiplying both sides by by $\frac{\mathrm{d} h(t)}{\mathrm{d} t}$ and integrating yields
\begin{equation}
	\frac{1}{2} \left(\frac{\mathrm{d} h(t)}{\mathrm{d} t}\right)^2 = \xi^2 d g \ln \left(\frac{d+h}{d}\right) - g h + C.
\end{equation}

Applying the initial conditions $h(t=0)=0$ and $\frac{\mathrm{d} h(t=0)}{\mathrm{d} t} = 0$ gives 
$C = 0$, and thus the first integral is
\begin{equation}
	\frac{1}{2} \left(\frac{\mathrm{d} h(t)}{\mathrm{d} t}\right)^2 = \xi^2 d g \ln \left(\frac{d+h}{d}\right) - g h,
\end{equation}
Therefore,
\begin{equation}
	\frac{\mathrm{d} h(t)}{\mathrm{d} t} = \sqrt{2 \left[\xi^2 d g \ln \left(\frac{d+h}{d}\right) - g h\right]},
\end{equation}
and the implicit time–height relation can be written as
\begin{equation}
	t(h) = \int_{0}^{h} \frac{\mathrm{d}h'}{\sqrt{2 \left[\xi^2 d g \ln \left(\frac{d+h}{d}\right) - g h\right]}},
\end{equation}

This integral has no closed-form elementary solution; therefore, we solved it numerically and inverted it to obtain 
$h(t)$.

Now, let us consider the case where the filament density is reduced. The force acting on the upper current can then be expressed as
\begin{equation}
	F = \frac{A}{d + h} - \frac{A}{\chi d}, 
\end{equation}
where $\chi$ denotes the factor by which the filament density is diluted. The corresponding mass is 
$m = \frac{A}{\chi dg}$. After substituting this into the equation of motion, we obtain 
\begin{equation}
	\frac{\mathrm{d}^2 h (t)}{\mathrm{d} t^2 } = \frac{\chi dg}{d+h} - g.
\end{equation}
As can be seen, this equation has the same form as in the case of increased current (Eq.~\ref{a4}), and thus it can be solved in the same manner. 

Note that in the both cases the filament trajectory does not depend on the electric current I and mass m.
As an example of the calculations, in Figure~\ref{graph1} we show the time evolution of the height $h$ and velocity $\rm v$ of the filament for the parameter $\xi = 4$, which corresponds to the $4\,B_0$ MHD case. Similarly, in Figure~\ref{graph2} we present the same for the parameter $\chi = 10$, which corresponds to the $0.1\,\rho_0$ MHD case. The both these figures show oscillations.

\end{appendix}

\end{document}